\documentclass[english,12pt]{article}
\usepackage{array}
\usepackage{graphicx}
\usepackage{amssymb}
\usepackage{amsmath}
\usepackage{multirow}
\usepackage{prettyref}
\usepackage{babel}
\usepackage{units}
\usepackage[latin1]{inputenc}
\usepackage{amsfonts}
\usepackage{amssymb}
\usepackage{babel}
\usepackage{color}
\usepackage{cite}
\def\@fmsl@sh#1#2#3{\m@th\ooalign{$\hfil#1\mkern#2/\hfil$\crcr$#1#3$}}
 \def\eq#1\en{\begin{equation}#1\end{equation}}
\def\s[#1,#2]{[#1\stackrel{\star}{,}#2]}
\def\sx[#1,#2]{[#1\stackrel{\star_{x}}{,}#2]}

\newcommand{\nc}{\newcommand}
\nc{\beq}{\begin{equation}}
\nc{\eeq}{\end{equation}}
\nc{\beqa}{\begin{eqnarray}}
\nc{\eeqa}{\end{eqnarray}}

\def\bc{\begin{center}}
\def\ec{\end{center}}

\def\gsim{\mathrel{\mathpalette\atversim>}}

\def\bc{\begin{center}}
\def\ec{\end{center}}

\def\gsim{\mathrel{\rlap{\lower4pt\hbox{\hskip1pt$\sim$}}

    \raise1pt\hbox{$>$}}}       

\def\gsim{\mathrel{\rlap{\lower4pt\hbox{\hskip1pt$\sim$}}
    \raise1pt\hbox{$>$}}}       

\begin{document}
\makeatletter
\def\fmslash{\@ifnextchar[{\fmsl@sh}{\fmsl@sh[0mu]}}
\def\fmsl@sh[#1]#2{%
  \mathchoice
    {\@fmsl@sh\displaystyle{#1}{#2}}%
    {\@fmsl@sh\textstyle{#1}{#2}}%
    {\@fmsl@sh\scriptstyle{#1}{#2}}%
    {\@fmsl@sh\scriptscriptstyle{#1}{#2}}}
\def\@fmsl@sh#1#2#3{\m@th\ooalign{$\hfil#1\mkern#2/\hfil$\crcr$#1#3$}}
\makeatother

\thispagestyle{empty}
\begin{titlepage}
\boldmath
\begin{center}
  \Large {\bf  Gravitational Effective Action at Second Order in Curvature and Gravitational Waves}
    \end{center}
\unboldmath
\vspace{0.2cm}
\begin{center}
{  {\large Xavier Calmet}\footnote{x.calmet@sussex.ac.uk}$^a$,  {\large Salvatore Capozziello}\footnote{capozziello@na.infn.it}$^{b,c,d}$  {\large and} {\large Daniel Pryer}\footnote{dp83@sussex.ac.uk}$^a$}
 \end{center}
\begin{center}
$^a${\sl Department of Physics $\&$ Astronomy, 
University of Sussex, Brighton, BN1 9QH, United Kingdom 
}\\
$^b${\sl  Dipartimento di Fisica ``E. Pancini", Universit\`a di Napoli ``Federico II'', Via Cinthia, I-80126, Napoli, Italy. }\\
$^c${\sl Istituto Nazionale di Fisica Nucleare (INFN), Sez. di Napoli, Via Cinthia 9, I-80126 Napoli, Italy.}\\
$^d${\sl Gran Sasso Science Institute, Via F. Crispi 7, I-67100, L'Aquila, Italy.}
\end{center}
\vspace{5cm}
\begin{abstract}
\noindent
We  consider the full effective theory for quantum gravity at second order in curvature including non-local terms. We show that the theory contains two new degrees of freedom beyond the massless graviton: namely a massive spin-2 ghost and a massive scalar field. Furthermore, we show that it is impossible to fine-tune the parameters of the effective action to eliminate completely the classical spin-2 ghost because of the non-local terms in the effective action. Being a classical field, it is not clear anyway that this ghost is problematic. It simply implies a repulsive contribution to Newton's potential.  We then consider how to extract the parameters of the effective action and show that it is possible to measure, at least in principle, the parameters of the local terms independently of each other using a combination of observations of gravitational waves and measurements performed by pendulum type experiments searching for deviations of Newton's potential.  
\end{abstract}  
\end{titlepage}



\newpage

\section{Introduction}

The aim of this work is to propose a new method to extract the parameters of the full effective field theory for quantum gravity working at quadratic order in curvature. Furthermore, we study the field content of this full effective field theory, i.e., including both local and non-local operators. We show that, beyond the massless spin-2 field, the effective theory contains a massive spin-2 field as well as a massive scalar field: the non-local operators do not introduce new fields beyond those identified by Stelle a long time ago \cite{Stelle:1977ry}. We point out that the mass of the spin-2 field can be bound by using recent gravitational wave observations \cite{Abbott:2016blz,Abbott:2016nmj,Abbott:2017vtc,laurentis}. This is interesting as the usual way to bound the masses of the scalar and spin-2 fields is to use measurements from the Cavendish pendulum type experiments (e.g. E\"ot-Wash) \cite{Hoyle:2004cw,Kapner:2006si,Tan:2016vwu} which are however sensitive to a combination of these masses.  A combination of the measurements of both Cavendish experiments and gravitational wave observations would, in principle, allow to disentangle the measurements of these two mass parameters. It is easy to understand why. While both massive fields contribute to a modification of Newton's potential, only the massive spin-2 can be excited by the quadrupole momentum of the merging two black holes system. We shall first briefly review the effective field approach to quantum gravity and then discuss our new proposal.

The quantization of general relativity remains one of the holy grails of theoretical physics. It is well understood that because Newton's constant is dimensionful,  general relativity is not renormalizable, at least not perturbatively. This incompatibility between classical general relativity and quantum field theory has been the motivation to study extensions of quantum field theory such as e.g. string theory which could incorporate a consistent theory of quantum gravity. Despite much work in this direction, it is still unclear what is the correct ultra-violet completion of Einstein's theory as making the connection to infra-red physics, and thus testable physics, is not straightforward. Problems are also present at infrared scales because retaining strictly Einstein's theory forces to introduce dark matter and dark energy at astrophysical and cosmological scales \cite{report,reportsergei,mauro,annalen}.

While we may be very far away from discovering the correct theory of quantum gravity, there is one framework which enables one to do quantum calculations within general relativity while remaining agnostic about physics above the reduced Planck scale $M_P$ or some $2.4 \times10^{18}$ GeV which is usually assumed to be the scale of quantum gravity. If one is interested in probing quantum gravity experimentally at energies below $M_P$, an effective field theory approach to quantum gravity, effective quantum gravity (EQG), is a possible approach (see e.g. \cite{Donoghue:1994dn,sergei,Calmet:2013hfa}). The effective theory approach is self-consistent up to the scale of quantum gravity $M_\star$, often identified with the reduced Planck mass. Remarkably, this approach enables model independent calculations in quantum gravity without having a full knowledge of the ultra-violet theory. It is sufficient to specify the symmetries of the theory below the Planck mass and the field content. Furthermore, some of the Wilson coefficients of EQG are calculable given this input while the remaining must be measured in experiments as we do not have the full ultra-violet theory to match these Wilson coefficients to the fundamental theory.

Assuming general coordinate invariance below the Planck scale, the effective field theory describing the dynamics for the metric $g_{\mu\nu}$ ( which is a massless spin-2 field), a cosmological constant $\Lambda_C$ and the standard model of particle physics $\mathcal{L}_{SM}$ (including the Higgs doublet $H$ ) is given by
\begin{eqnarray}\label{action1}
S &=& \int d^4x \, \sqrt{-g} \left[ \left( \frac{1}{2}  M^2 + \xi H^\dagger H \right)  \mathcal{R}- \Lambda_C^4 + c_1 \mathcal{R}^2 + c_2 \mathcal{R}_{\mu\nu}\mathcal{R}^{\mu\nu}+ c_4   \Box \mathcal{R}  \right . \nonumber \\
&& \left . + b_1 \mathcal{R} \log \frac{\Box}{\mu^2_1}\mathcal{R} + b_2 \mathcal{R}_{\mu\nu}  \log \frac{\Box}{\mu^2_2}\mathcal{R}^{\mu\nu}  
+ b_3 \mathcal{R}_{\mu\nu\rho\sigma}  \log \frac{\Box}{\mu^2_3}\mathcal{R}^{\mu\nu\rho\sigma} + \mathcal{O}(M_\star^{-2}) + \mathcal{L}_{SM}  \right], 
\end{eqnarray}
where $\mathcal{R}$, $\mathcal{R}^{\mu\nu}$ and $\mathcal{R}^{\mu\nu\rho\sigma}$ are respectively the Ricci scalar, Ricci tensor and Riemann tensor  and  $\mu_i$ are renormalization scales.   Note that each of these structures are functions of the metric and they contain second order derivatives. The effective action can be seen as a derivative expansion, in full analogy to chiral perturbation theory in quantum chromodynamics. It is obtained by integrating out the graviton and massless matter fields (see e.g. \cite{Donoghue:2014yha}). The calculation is done at the one-loop level in perturbation theory using dimensional regularization, the divergencies of the diagrams giving rise to the non-local terms of the type $\mathcal{R} \log \Box \mathcal{R}$ are absorbed in the corresponding local terms $\mathcal{R}^2$ and $\mathcal{R}_{\mu\nu}\mathcal{R}^{\mu\nu}$. In the following, we drop the total derivative $\Box \mathcal{R}$ as it does not affect the equations of motion. Note that the Riemann tensor squared term $\mathcal{R}_{\mu\nu \alpha \beta}\mathcal{R}^{\mu\nu \alpha \beta} $ can be eliminated using the Gauss-Bonnet identity: this cannot be done though for the corresponding non-local term. It is worth emphasizing that the effective action could be constrained further if we imposed new symmetries such as conformal invariance, see e.g. \cite{Antoniadis:1992xu,Odintsov:1991tg}, here we choose to stick to Einstein's formulation of gravity as the leading order term of our effective action. We shall now describe the parameters of this effective action and describe its dynamical content.

\section{The parameters of the effective action and its dynamical content}

The effective action contains both dimensionful and dimensionless parameters. The most familiar one is certainly the reduced Planck scale $M_P$ which is given by
\begin{eqnarray}
\label{effPlanck}
M_P^2=(M^2+\xi v^2) \, ,
\end{eqnarray}
where $v=246$ GeV is the Higgs boson's expectation value and $\xi$ is the non-minimal coupling of the Higgs boson. The non-minimal coupling is a free parameter unless conformal invariance is imposed. Measurements of the properties of the Higgs boson imply that  $|\xi| >2.6 \times 10^{15}$ is excluded at the $95 \%$ C.L.  \cite{Atkins:2012yn}. $M$ is the coefficient of the Ricci scalar. It has mass dimension 2. The scale $M_\star$ is the scale up to which we can trust the effective field theory. It is traditionally identified with $M_P$ but this needs not to be the case.  Direct searches for strong gravitational effects at colliders in the form of quantum black holes \cite{Calmet:2008dg} lead to a bound on $M_\star$ of the order of 9 TeV, see e.g. \cite{Sirunyan:2017anm}. The renormalization scales $\mu_i$ could, in principle, be different for the three non-local operators, but we will assume that $\mu_i=\mu$. It seems reasonable to take it of the order of $M_\star$ as this is the energy scale at which the effective theory needs to be matched to the underlying theory of quantum gravity. 

While the Wilson coefficients of the local operators $\mathcal{R}^2$ and $\mathcal{R}_{\mu\nu}\mathcal{R}^{\mu\nu}$ are not calculable within the effective field theory approach, the Wilson coefficients $b_i$ of the non-local operators are calculable from first principles and are truly model independent predictions of quantum gravity. Their values are reproduced in  Table \ref{table1}. 
\begin{table}[h!]
\begin{center} 
\begin{tabular}{ |c|c|c|c|}
 \hline
   & $b_1$ & $b_2$ & $b_3$ \\
  \hline 
real scalar& $5(6 \xi -1)^2/(11520 \pi^2)$ &  $-2/(11520 \pi^2)$& $2/(11520 \pi^2) $ \\ 
 \hline
Dirac spinor& $-5/(11520 \pi^2)$ & $ 8/(11520 \pi^2)$ & $7/(11520 \pi^2)$  \\ 
 \hline
 vector & $ -50/(11520 \pi^2)$ & $176/(11520 \pi^2)$ & $ -26/(11520 \pi^2)$ \\ 
 \hline
graviton & $430/(11520 \pi^2)$ & $-1444/(11520 \pi^2)$ & $424/(11520 \pi^2)$ \\ 
 \hline 
\end{tabular}
\caption{Calculable Wilson coefficients, see e.g. \cite{Donoghue:2014yha} where they are calculated using dim-reg. These results match the classical ones published in \cite{Birrell:1982ix} where they are calculated using the conformal anomaly method.}
\label{table1}
\end{center}
\end{table}
The effective action can be linearized around flat space-time. One obtains
		\begin{align}	\nonumber 
	\Box \left[h_{\mu\nu} - \frac{1}{2} \eta_{\mu\nu} h \right] +
	 \kappa^2 &\bigg[ 
	 \left [ \left(b_1 + \frac{b_2}{4}\right)\log\left(\frac{\Box}{\mu^2}\right)+ \left (c_1 + \frac{c_2}{4}\right ) \right]  \eta_{\mu\nu}   \Box^2  h  \\ \nonumber
	& 	  - \left [  \left( b_1 + \frac{b_2}{2} + b_3 \right)  \log\left(\frac{\Box}{\mu^2}\right) +
	 \left( c_1 + \frac{c_2}{2}\right) \right ] \partial_\mu \partial_\nu \Box h \\
	&+ \left [ \left(\frac{b_2}{2} + 2 b_3 \right) \log\left(\frac{\Box}{\mu^2}\right) +\frac{c_2}{2}  \right ] \Box^2 h_{\mu\nu} \bigg] = 0 \ \ ,
	\end{align}	
where we used the harmonic gauge ( $\partial_\nu h^{\mu\nu} = \frac{1}{2}\partial^\mu h$) and $\kappa^2 = 32\pi G$. It is straightforward to see that the effective action contains two new degrees of freedom besides the massless spin-2 ``classical'' graviton (the ``quantum graviton'' has been integrated out of the effective action). We have a massive spin-2 field and a massive scalar field. The linearized effective action reads
\begin{eqnarray}
S &=& \int d^4x \left \{ -\frac{1}{4} h^{\mu\nu} \left [- \left(c_2+\left (b_2+ 4 b_3 \right ) \log\left(\frac{\Box}{\mu^2}\right) \right ) \kappa^2 \Box+2 \right ] \Box P^{(2)}_{\mu\nu\rho\sigma} h^{\rho\sigma}  \right .  \nonumber \\ && \left .
+\frac{1}{2} h^{\mu\nu} \left [ 2\left (3 c_1 + c_2+\left(3 b_1+ b_2+ b_3 \right)\log\left(\frac{\Box}{\mu^2}\right) \right ) \kappa^2 \Box+2\right ]\Box P^{(0)}_{\mu\nu\rho\sigma} h^{\rho\sigma}  
+\kappa h_{\mu\nu} T^{\mu\nu}\right \} \nonumber \\ &&
\end{eqnarray}
with
\begin{align}
P^{(2)}_{\mu\nu\rho\sigma} = \frac{1}{2} (L_{\mu\rho} L_{\nu\sigma}+L_{\mu\sigma} L_{\nu\rho}) -\frac{1}{3}L_{\mu\nu} L_{\rho\sigma},
\end{align}
\begin{align}
P^{(0)}_{\mu\nu\rho\sigma} = \frac{1}{3}L_{\mu\nu} L_{\rho\sigma},
\end{align}
where $L_{\mu\nu}=\eta_{\mu\nu} -\partial_\mu \partial_\nu/\Box$.

The dynamical content of the theory can be made explicit by calculating
\begin{eqnarray}
&&T^{(1)\mu\nu}  \left [ -\frac{1}{4}  \left (- \left(c_2+\left (b_2+ 4 b_3 \right ) \log\left(\frac{\Box}{\mu^2}\right) \right ) \kappa^2 \Box+2 \right ) \Box P^{(2)}_{\mu\nu\rho\sigma}   \right .   \nonumber \\ && \left . 
+\frac{1}{2}  \left ( 2\left (3 c_1 + c_2+\left(3 b_1+ b_2+ b_3 \right)\log\left(\frac{\Box}{\mu^2}\right) \right ) \kappa^2 \Box+2\right )\Box P^{(0)}_{\mu\nu\rho\sigma} \right ] T^{(2)\rho\sigma}  
\end{eqnarray}
where $T^{(1)\mu\nu}$ and $T^{(2)\rho\sigma}$ are two conserved sources. In momentum space, one obtains:
\begin{eqnarray} \label{fc}
\frac{\kappa^2}{4} \left [ \frac{T^{(1)}_{\mu\nu} T^{(2)\mu\nu} - \frac{1}{2} T^{(1)\mu}_{\ \ \mu} T^{(2)\nu}_{\ \ \nu}}{k^2}
-\frac{T^{(1)}_{\mu\nu} T^{(2)\mu\nu} - \frac{1}{3} T^{(1)\mu}_{\ \ \mu} T^{(2)\nu}_{\ \ \nu}}{k^2-\frac{2}{ \kappa^2 \left(c_2+\left (b_2+ 4 b_3 \right ) \log \left( \frac{-k^2}{\mu^2} \right )\right)}}
+ \frac{T^{(1)\mu}_{\ \ \mu} T^{(2)\nu}_{\ \ \nu}}{k^2-\frac{1}{\kappa^2 \left (3 c_1 + c_2+\left(3 b_1+ b_2+ b_3 \right)\log\left(\frac{-k^2}{\mu^2}\right) \right )}}
 \right ]
\end{eqnarray}
to leading order in $\kappa^2$.
As mentioned before, the effective action contains, besides the usual massless graviton (first term in Eq. (\ref{fc})), a massive spin 2 particle (second term in Eq. (\ref{fc})) and a massive scalar field (third term in Eq. (\ref{fc})). Because of the negative sign in front of the second term, the massive spin 2 object carries negative energy, i.e., it is a ghost. It should however be kept in mind that we are considering the effective action obtained after integrating out the particles. This ghost thus does not need to be quantized and it is a classical field. 
The mass of the spin-2 ghost is given by the solution to the equation
\begin{eqnarray}
k^2-\frac{2}{ \kappa^2 \left(c_2+\left (b_2+ 4 b_3 \right ) \log \left( \frac{-k^2}{\mu^2} \right )\right)}=0.
\end{eqnarray}
One finds
\begin{align}
m^2=\frac{2}{ (b_2+ 4 b_3) \kappa^2 W\left(-\frac{2 \exp\frac{c_2}{(b_2+ 4 b_3)}}{ (b_2+ 4 b_3) \kappa^2 \mu^2}\right)},
\end{align}
where $W(x)$ is the Lambert function. The squared mass is, in general, a complex number and the pair of complex ghosts will thus have a width with an extremely short lifetime close to the Planck time \cite{Calmet:2014gya}. The conservative assumption is that the presence of these poles simply signals a breakdown of perturbation theory at the corresponding energy scale. This is the true scale of quantum gravity $M_\star$ and the effective field theory must be abandoned at this energy scale.

One may be tempted to shift the mass of the ghost above the reduced Planck mass to extend the validity range of the effective field theory by adjusting the coefficient $c_2$ to be very small or zero as it is sometimes advocated \cite{Stelle:1977ry}.  However, it is clear that setting $c_2=0$ will not remove the ghost.  The non-local terms will not be eliminated by this choice and as emphasized before, the Wilson coefficients of the non-local terms are not free parameters but rather they are calculated from first principles. A small $c_2$ would not compensate the contribution from the non-local term. Let us introduce the parameter 
\begin{equation}
N=\frac{1}{6}\left[N_S(b_2+ 4 b_3)_{scalar}+N_F(b_2+ 4 b_3)_{fermion}+N_V(b_2+ 4 b_3)_{vector}\right]\,,
\end{equation}
 where $N_S$, $N_F$ and $N_V$ are respectively the number of real scalars, Dirac fermions and real vector fields in the theory. We see that $N$ cannot be too large or the mass of the ghost will drop below the reduced Planck mass and we would have to abandon the effective theory below the reduced Planck mass. These results are in accordance with previous works \cite{Calmet:2014gya,Calmet:2015pea,Calmet:2017omb} where it was shown that although perturbative unitarity can be restored below the Planck mass, the presence of ghosts signals the breakdown of the effective effective field theory. It was shown in \cite{Calmet:2014gya}, that this energy scale is associated with strong quantum gravitational effects. 

 Furthermore, it is impossible to find a combination of matter fields that would compensate the graviton contribution to $b_2+4 b_3$ which can be written as $252 + 36 N_f + 6 N_s + 72 N_v$. This quantity is positive and larger than 252 for any matter content. There is thus no obvious manner to avoid the massive spin-2 ghost. We would also like to point out that setting $c_2=0$ is not very satisfactory anyway. It is a renormalized coupling constant (see e.g. \cite{Elizalde:1995tx}), while it may take the value 0 at some energy scale, it would take some symmetry argument to enforce $c_2(\mu)=0$ at all scales. Note that the physical consequences of the renormalization group equations for the coefficients of the local part of the action have been investigated in \cite{Elizalde:2017mrn,Myrzakulov:2016tsz,Myrzakulov:2014hca}. Obviously, the renormalization group equation of $c_2$ modifies the structure of the massive ghost at higher order in perturbation theory.

We thus have to accept that the effective field theory, which contains classical fields after the quantum fields have been integrated out, contains a classical ghost. Whether or not this is a problem remains to be understood, however it does not appear to be a dramatic issue as a classical ghost may not cause any instability in contrast to quantum ghosts. 
Let us now turn out attention to experimental bounds on the coefficients of the effective field theory to second order in curvature.

\section{How to measure the masses of the massive spin-2 and spin-0 fields?}

We have seen that the Wilson coefficients $b_i$ are small unless the number of fields $N$ introduced in the model is very large. On the other hand, $c_i$ could be arbitrarily large and we shall thus assume that they are larger than $b_i$. Unless $c_i$ are very large (as we shall see at least $10^{61}$), then their effect on any observable is minuscule. Our approximation is thus certainly a good one.  Stelle \cite{Stelle:1977ry} has pointed out that it is possible to derive bounds on $c_1$ and $c_2$ using  torsion pendulum experiments of the  Cavendish type experiments searching for deviations of Newton's potential. Indeed,the terms  $c_1 \mathcal{R}^2$ and $c_2 \mathcal{R}^{\mu\nu}\mathcal{R}_{\mu\nu}$ lead to Yukawa-like corrections to Newton's potential of a point mass $m$: 
\begin{eqnarray}
\Phi(r) = -\frac{Gm}{r} \left( 1+\frac{1}{3}e^{-m_0 r}-\frac{4}{3}e^{-m_2 r} \right) 
\end{eqnarray}
with
\begin{eqnarray}
m_0^{\,-1}=\sqrt{32\pi G \left(3c_1-c_2 \right)}
\end{eqnarray}
and
\begin{eqnarray}
m_2^{\,-1}=\sqrt{16\pi G c_2}.
\end{eqnarray}
Sub-millimeter tests of Newton's law \cite{Hoyle:2004cw} using sophisticated pendulums are used to bound $c_1$ and $c_2$. One finds that,  in the absence of accidental fine cancellations between both Yukawa terms, they are  constrained to be less than $10^{61}$ \cite{Hoyle:2004cw,Calmet:2008tn} . Note that this bound is obtained assuming that the $b_i$ Wilson coefficients are small which as argued above is a reasonable assumption.

Torsion pendulum experiments do not allow to measure $c_1$ and $c_2$ independently.  Recently, it was argued that is it possible to bound $c_1$ using data from black hole mergers on $c_1$ \cite{Cao:2013osa}.  However, this bound was obtained under the assumption that  there is a shell of  scalar field around the binary black hole which is not a necessity. Here we want to point out another possibility, to measure $c_1$ and $c_2$ independently.  This involves using the recent discovery of gravitational wave measurements on top of the usual Cavendish type experiment\footnote{We note that upper bounds on $m_0$ and $m_2$ can be achieved also by satellite experiments \cite{mairi} or stellar dynamics around the Galactic Centre \cite{borka}.}.

The detections of three binary black hole mergers were recently announced  \cite{Abbott:2016blz,Abbott:2016nmj,Abbott:2017vtc}. In each case, the final merged black hole had a smaller mass than the two progenitors, with this lost mass being radiated away as gravitational waves. The LIGO-VIRGO collaboration has used their observations to set a limit on the mass of the graviton. Constraining the graviton mass from these observations is performed by comparing arrival times of the signals at each detector; a massive graviton implies subluminal velocity, with differing frequencies of gravitational waves propagating at different speeds. The bound on a graviton mass derived by LIGO-VIRGO reads \cite{Abbott:2016blz}
\begin{eqnarray}
m_g  < 1.2 \times 10^{-22} \mbox{eV}.
\end{eqnarray}

To obtain a conservative bound on $c_2$, we assume that all of the energy of the merger is emitted into the massive spin-2 field.
The Wilson coefficient $c_2$ is related to mass $m_g$ via
\begin{eqnarray} 
c_2= \frac{1}{16 \pi G m_g^2}.
\end{eqnarray}
We thus get the following bound
\begin{eqnarray}
c_2 < \frac{\left(1.22 \times 10^{28}\mbox{eV}\right)^2}{16\pi \left(1.2 \times 10^{-22}\mbox{eV}\right)^2} = 2.1\times 10^{98} .
\end{eqnarray}
Although this is not a strong bound, it does, however, directly apply to the coefficient of $\mathcal{R}^{\mu\nu}\mathcal{R}_{\mu\nu}$ and it is independent of the $\mathcal{R}^2$ term. 

One may wonder whether scalar gravitational waves might be produced during the merger of the two black holes. However, as shown in \cite{Thorne:1980ru} for a Kerr metric, the mass monopole represents the total mass-energy in a system, which is conserved thus it does not give off radiation. This implies that the scalar field, present in the classical propagator,  cannot be produced. The mass dipole corresponds to the center of mass of the system.  Its first derivative is the angular momentum which is conserved as well. On the other hand, the mass quadrupole moment is not constant in time and it is the source of the emission of spin-2 gravitational waves. Both the massive and massless spin-2 fields can be produced. We note that the local and non-local terms at second order in curvature do not modify the Schwarzschild metric at this order in the curvature expansion \cite{Calmet:2017qqa}. The same applies to the Kerr metric. Furthermore, due to their design as interferometers, gravitational wave detectors are only sensitive to the changes of the quadrupole moment, i.e., to waves of spin-2.

Thus far we have assumed that only the massive spin-2 wave was produced to derive our bound. Improving the bound and measuring directly $c_2$ requires to differentiate between the massive and massless spin-2 modes. The massless spin-2 mode has two polarizations while the massive mode has five, see e.g. \cite{vanDam:1970vg}. It has been shown in \cite{Bogdanos:2009tn,Capozziello:2015nga} that the two modes, because of the differences in their polarizations, would lead to different signatures in the interferometer detectors  such as the LIGO-VIRGO system and it is thus possible to disentangle these two contributions. This is the final ingredient necessary to establish that $c_2$ can, in principle, be measured independently of $c_1$ using gravitational wave detectors.

\section{Conclusions}

In this paper, we have considered the full effective theory for quantum gravity at second order in curvature including non-local terms. We have shown that the theory contains two new degrees of freedom beyond the massless graviton namely a massive spin-2 ghost and a massive scalar field. We have shown that it is impossible to fine-tune the parameters of the effective action to eliminate completely the classical spin-2 ghost. As this is a classical field, it is not clear anyway that this ghost is problematic as it is not obvious that it would lead to physical pathologies. It simply implies a repulsive contribution to Newton's potential.  We then have considered how to extract the parameters of the effective action and shown that it is possible to measure, at least in principle, the parameters of the local terms independently of each other using a combination of observations of gravitational waves and measurements performed using pendulum type experiments searching for deviations of Newton's potential.  

{\it Acknowledgments:}
The work of XC is supported in part  by the Science and Technology Facilities Council (grant number  ST/J000477/1). 
SC acknowledges the COST Action CA15117 (CANTATA)
and INFN Sez. di Napoli (Iniziative Specifiche QGSKY
and TEONGRAV). 


\bigskip{}

\end{document}